\documentclass[acmtog, nonacm]{acmart}

\usepackage{amsmath}
\usepackage{mathtools}
\usepackage{bm}
\usepackage{graphicx}
\usepackage[export]{adjustbox}
\usepackage[percent]{overpic}
\usepackage{wrapfig}
\usepackage{rotating}
\usepackage{units} 
\usepackage{tabularx}
\usepackage{makecell}
\usepackage{booktabs} 
\usepackage{algpseudocode}
\usepackage{tabularray}
\usepackage{multirow}
\usepackage[ruled]{algorithm2e}

\usepackage{subfig} 
\DeclareMathOperator*{\argmin}{argmin}
\usepackage[dvipsnames,table,xcdraw]{xcolor} 

\usepackage{tikz}
\usepackage{tikzscale}
\usepackage{pgfplots}
\pgfplotsset{compat=1.18}
\usetikzlibrary{
  backgrounds,
  calc,
  fit,
  spy,
  decorations.pathreplacing,
  arrows.meta,
  tikzmark,
  shapes.geometric
}

\tikzset{every node/.style={font=\fontfamily{LinuxBiolinumT-TLF}\selectfont}}

\usepackage{cleveref}

\newlength{\subcolumnwidth}

\newcommand{\nextsubcolumn}[1][]{%
  \cr\noalign{\hfill}
  \if\relax\detokenize{#1}\relax\else\hsize=#1\setlength{\subcolumnwidth}{\hsize}\fi
}

\SetAlFnt{\small}
\SetAlCapFnt{\small}
\SetAlCapNameFnt{\small}
\SetAlCapHSkip{0pt}

\makeatletter
\newcommand{\showfontinfo}{
    Font: \f@family \f@series \f@shape, Size: \f@size
}
\makeatother
\begin{document}
\title{Odd-DC: Generalizable Neural Model Reduction \\ via Odd Difference-of-Convex Structure}

\author{Towa Shixun Huang}
\orcid{TBD}
\affiliation{%
  \institution{University of Toronto}
  \country{Canada}}
\email{towaxun.huang@mail.utoronto.ca}

\author{Eitan Grinspun}
\orcid{0000-0003-4460-7747}
\affiliation{%
  \institution{ University of Toronto}
  \country{Canada}}
\email{eitan@cs.toronto.edu}

\author{Yue Chang}
\orcid{0000-0002-2587-827X}
\affiliation{%
  \institution{University of Toronto}
  \country{Canada}}
\email{changyue.chang@mail.utoronto.ca}

\begin{abstract}

Model reduction is essential for real-time simulation of deformable objects. Linear techniques such as PCA provide structured and predictable behavior, but their limited expressiveness restricts accuracy under large or nonlinear deformations. Nonlinear model reduction with neural networks offers richer representations and higher compression; however, without structural constraints, the learned mapping from latent coordinates to displacements often generalizes poorly beyond the training distribution.

We present an odd difference-of-convex (DC) neural formulation that bridges linear and nonlinear model reduction. Our goal is to obtain a latent space that behaves reliably under unseen load magnitudes and directions. To improve extrapolation in magnitude, we introduce convexity into the decoder to discourage oscillatory responses. Yet convexity alone cannot represent the odd symmetry required by many symmetric systems, which is crucial for generalization to inverse force directions. We therefore adopt a DC formulation that preserves the stabilizing effect of convexity while explicitly enforcing odd symmetry. Practically, we realize this structure using an input-convex neural network (ICNN) augmented with symmetry constraints.

Across challenging deformation scenarios with varying magnitudes and reversed load directions, our method demonstrates stronger generalization than unconstrained nonlinear reductions while maintaining compact latent spaces and real-time performance. Our DC formulation extends to both mesh-based and neural-field reductions, demonstrating applicability across multiple classes of neural nonlinear model reduction.
    
\end{abstract}

%
%


%
%



\begin{teaserfigure}
    \centering
    \includegraphics[width=\linewidth]{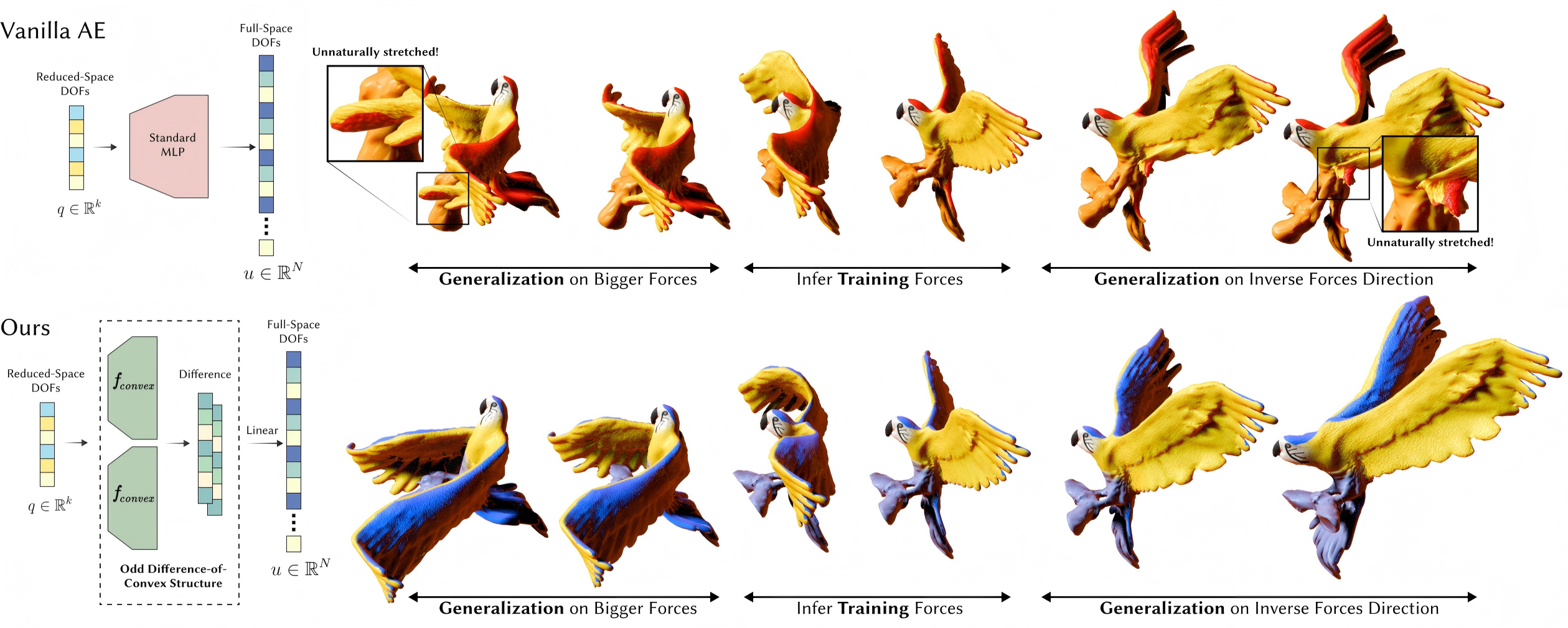}
    \caption{We regularize nonlinear model reduction through a convex-inspired functional design. 
The convex structure discourages oscillations in the displacement field, while an odd-symmetric construction captures the intrinsic symmetry of physical deformation. 
Together, these constraints yield smoothly varying displacements in the latent reduced space, even for coordinates outside the training set. 
We visualize the displacement norm and the corresponding full-space deformations mapped from points in the latent reduced space. }
    \label{fig:TEASER}
\end{teaserfigure}

\maketitle

\section{Introduction}
\label{sec:introduction}

Real-time simulation of deformable objects has long been a central topic in computer graphics. Among the various acceleration strategies, model reduction remains one of the most effective, often yielding orders-of-magnitude speedups while retaining physical fidelity.

A widely adopted paradigm for model reduction is data-driven method. Early approaches~\cite{barbivc2005real} collected snapshot data from full-space simulations and constructed a linear subspace via principal component analysis (PCA). This linear approximation enables fast runtime evaluation but limits expressiveness due to the inherent linearity of the representation. To address this, Fulton et~al.~\cite{Fulton:LSD:2018} introduced a nonlinear model reduction framework that learns an additional mapping from a low-dimensional latent space to the PCA-reduced coordinates. This nonlinear formulation achieves greater compression, faster convergence, and improved online performance compared to its linear counterpart.

While both linear and nonlinear data-driven methods perform well within the latent space spanned by their training data, their behavior beyond this range differs fundamentally. For linear bases such as PCA, the linearity of the mapping enforces a structured form of generalization: when the latent coordinates move outside the training region, the corresponding deformations scale consistently along those directions. In contrast, nonlinear model reduction—typically parameterized by neural networks—lacks such structural constraints. Outside the training distribution, the displacements can behave unpredictably, often producing implausible or unstable deformations. This exposes a fundamental dilemma: linear models offer structured predictability but limited expressiveness, whereas nonlinear models provide flexibility at the cost of generalization. Our work bridges this divide by introducing a formulation that preserves the structural consistency characteristic of linear models while retaining the expressive power of nonlinear ones.

In nonlinear model reduction, the mapping from reduced coordinates to full-space displacements remains highly expressive. The key question is therefore what structure should this mapping possess to generalize beyond the training set. Our goal is robustness to unseen load magnitudes and directions. To promote reliable extrapolation in magnitude, we seek a latent-to-displacement map whose response grows coherently rather than oscillating, motivating the introduction of convexity as a structural prior. Unlike loss-based regularizers that act only on sampled data, convexity constrains the mapping globally, including unsampled regions of the latent space that are inevitably visited during extrapolation.

However, convexity alone is insufficient for many approximately symmetric mechanical systems. In a wide range of elastic settings, loads applied in opposite directions tend to induce approximately opposite displacements. A purely convex mapping cannot automatically express such behavior, which can limit extrapolation when forces change sign. We therefore introduce an odd-symmetric difference-of-convex (DC) formulation, inspired by classical DC theory \cite{Hartman1959DC}, which enables antisymmetric responses while retaining the stabilizing role of convexity. This structure targets generalization in both magnitude (through convexity) and direction (through approximate odd symmetry).

To realize this DC construction in practice, we implement each convex component with an input-convex neural network (ICNN) \cite{amos2017icnn}, ensuring convexity with respect to the latent coordinates by design. We then induce odd symmetry by evaluating the network at a latent code and at its sign-reversed counterpart and combining the results in an antisymmetric manner. This strategy preserves the convex structure of the trainable network while producing a decoder that responds with opposite displacements to opposite latent directions.

We evaluate our method by focusing on generalization beyond the training distribution. Specifically, we test under external forces that differ substantially from those seen during training, including larger magnitudes and reversed load directions across more than fifty deformation scenarios. In these challenging settings, our approach demonstrates stronger generalization than unconstrained nonlinear reductions, producing qualitatively more realistic deformations and quantitatively improved accuracy while maintaining compact latent spaces and real-time performance.

Because our formulation remains nonlinear yet structured, it enables higher compression: accurate behavior can be achieved with fewer reduced dimensions, improving robustness when only a small number of cubature points are available. The proposed DC structure is also representation-agnostic. We show that it can be integrated into both mesh-based nonlinear reductions and continuous neural-field–based reductions, demonstrating applicability across multiple classes of neural model reduction while supporting real-time, interactive deformation.

In summary, we proposed
\begin{itemize}

\item \textbf{A Difference-of-Convex structural prior for nonlinear model reduction.}  
We propose an odd-symmetric DC decoder that improves extrapolation to unseen magnitudes and inverse directions while enabling higher compression and robust cubature, validated on 50+ mesh-based scenarios.

\item \textbf{Extension to continuous model reductions.}  
We show that the same DC prior regularizes continuous ROMs, outperforming prior discretization-agnostic nonlinear methods and enabling training and reconstruction from real-world data with generalization.

\end{itemize}

\begin{figure}
    \centering
    \includegraphics[width=1.\linewidth]{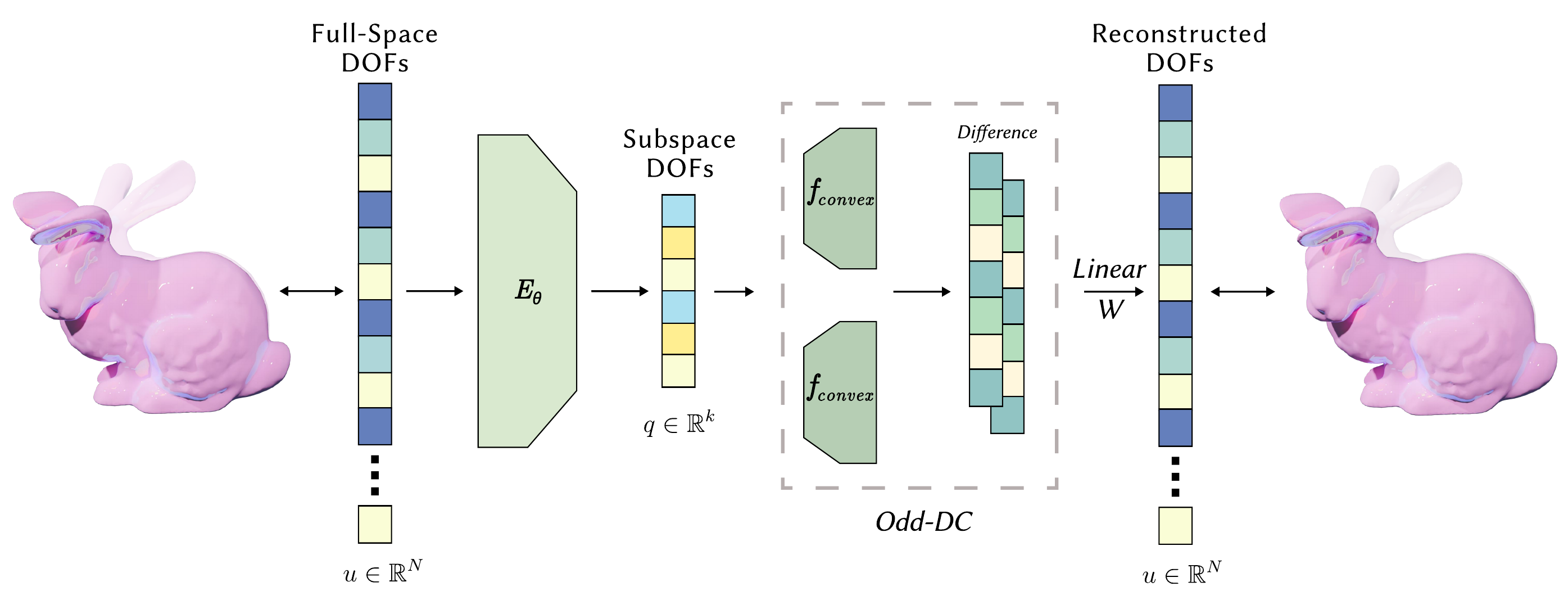}
    \caption{Overview of our construction. 
During training, the full-space deformation is first projected to a reduced space via an encoder. 
To reconstruct the deformation, we compute a nonlinear mapping from the reduced-space coordinates to an intermediate variable using an input-convex functional. 
This variable, of slightly higher dimensionality, is then augmented with an odd-symmetric construction to capture the intrinsic symmetry of the physical system. 
Finally, the augmented variable is linearly mapped back to the full-space displacement through the last linear layer. }
    \label{fig:overview}
\end{figure}

\begin{figure*}
    \centering
    \includegraphics[width=1.025\linewidth]{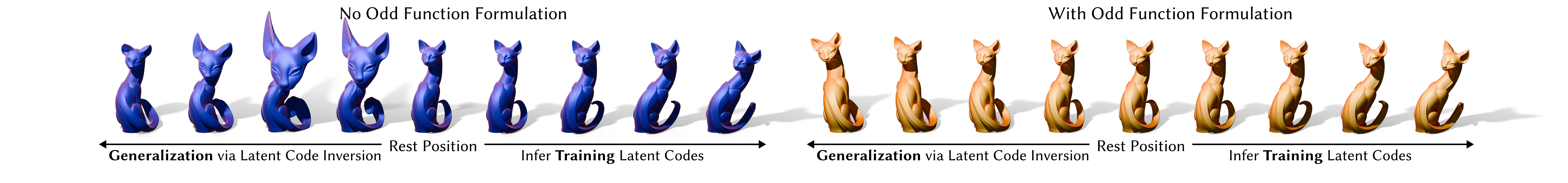}
    \caption{Effect of our odd-symmetric construction. 
Our odd-function formulation models the intrinsic symmetry of the system, enabling better generalization to deformations with opposite displacements. 
We train the model on a single deformation trajectory and visualize the resulting displacements from both the training latent codes and their inverses. 
Without the odd-function term (left), the model produces plausible deformations for training codes but exhibits higher variance and artifacts under latent inversion. 
With the odd-symmetric formulation (right), the deformations remain consistent across both original and inverted latent codes, demonstrating improved symmetry and generalization.
}
    \label{fig:odd_illus}
\end{figure*}

\section{Related Work}

\subsection{Reduced-space Simulation using Neural Networks}
 
Neural networks have become a popular tool for accelerating reduced-order modeling in physics-based simulation, with applications spanning fluids~\cite{deng2023neural, 10.1145/3641519.3657438, kim2019deep, tao2024neural}, contact and collisions~\cite{Cai:2022:CSDF, Romero:LCCHSD:2021, yang2020learning}, and deformable materials~\cite{CHEN:CROM-MPM:2023, Feng:2024:NAH, chen2023crom}. While prior work has explored both linear and nonlinear formulations—such as learning linear subspaces with neural priors~\cite{chang:2023:licrom, Modi:2024:Simplicits}—we specifically focus on nonlinear model reduction for deformable objects in this paper, as it forms the core of our method.

While the nonlinear expressivity of neural networks enables rich subspace representations~\cite{Fulton:LSD:2018, zong2023neural, Sharp:2023:datafree, shen2021high}, it often introduces overfitting and poor generalization beyond the training domain. Additionally, neural latent spaces typically require regularization to be more suitable for simulation. Common strategies include using a linear (PCA) layer~\cite{Fulton:LSD:2018, shen2021high}, adding soft penalties~\cite{Sharp:2023:datafree}, or encouraging a bounded Hessian Lipschitz constant~\cite{lyu2024accelerate}.

Our goal is similar in that we also seek to regularize the learned latent space. Like previous methods, ours can be combined with a PCA layer. However, instead of relying on additional loss terms~\cite{Sharp:2023:datafree, lyu2024accelerate}, we adopt a network architecture built from convex neural components arranged in a structured Difference-of-Convex formulation, embedding regularization directly into the model design.

\subsection{Regularization of Neural Networks}

Although neural networks offer rich representational power in input-output learning, that same capacity makes them prone to overfitting—achieving near-perfect fits on training deformations but struggling to generalize to novel configurations. In conventional settings, this is typically mitigated by augmenting the loss with weight‐decay $l_2$ penalties~\cite{ridge}, Lipschitz constraints via spectral‐norm regularization or Hessian‐Lipschitz terms~\cite{terjék2020adversariallipschitzregularization}, and Jacobian penalties that directly bound gradient magnitudes~\cite{Jacobson:2018:GWN}.  Such loss‐based techniques provide flexible control over network smoothness, but require careful tuning of hyperparameters and incur additional computational cost during training.

In reduced-space simulation, decoder overfitting can lead to nonphysical artifacts, slow convergence during energy minimization, and poor generalization to out-of-distribution inputs. Previous approaches have attempted to mitigate these issues through loss-based regularization~\cite{Sharp:2023:datafree, lyu2024accelerate}, but these methods rely on explicit penalty terms. This strategy has two key limitations. First, regularization is only effective if the loss is sufficiently optimized, making it difficult to determine the optimal training epoch. Second, these approaches require sampling the latent space; however, because the latent space is typically unbounded, it is impossible to sample exhaustively. As a result, the desired regularization properties may not hold for latent codes outside the sampled region. In contrast, our formulation enforces convexity in the neural network, providing stronger and more reliable guarantees. 

We enforce regularization architecturally via a specially designed Difference-of-Convex function, inspired by classical DC theory \cite{Hartman1959DC}. By constructing the decoder as a structured combination of convex components, we constrain its curvature and suppress highly oscillatory behavior without relying on extra penalty terms. This architectural constraint provides global regularization guarantees that hold across the entire latent space. Moreover, when combined with symmetric construction, the DC formulation naturally supports antisymmetric mappings, further enlarging the regime of reliable generalization.


\section{Background: Nonlinear Model Reduction and Generalization}
\label{sec:method}

\subsection{Nonlinear Model Reduction}
Let $ \mathbf{x}_0 \in \mathbb{R}^N $ denote the rest configuration of a discretized mesh with \( n \) nodes and \( m \) elements, where $ N = d n$ with $ d \in \{2, 3\} $ denoting the spatial dimension of the system. At time \( t \), the deformation is described by a displacement vector \( \mathbf{u} \in \mathbb{R}^N \), such that the current vertex positions are given by
\[
\mathbf{x}(t) = \mathbf{x}_0 + \mathbf{u}.
\]
The system evolves under Newton’s second law at time $t$:
\begin{equation}\label{eq:newton_rewrite}
\mathbf{M} \ddot{\mathbf{u}} + \nabla E(\mathbf{x}_0 + \mathbf{u}) = \bm{f}_{\mathrm{ext}}(t),
\end{equation}
where \( \mathbf{M} \in \mathbb{R}^{N \times N} \) is the lumped mass matrix, \( E \) denotes the elastic potential energy, and \( \bm{f}_{\mathrm{ext}}(t) \) is the external force (e.g., gravity, friction, or user interaction) applied at time \( t \).

Using implicit Euler integration, this second-order system is reformulated into a sequence of optimization problems over displacements. Given velocity \( \dot{\mathbf{u}}^t \) and position \( \mathbf{u}^t \), the next-step displacement \( \mathbf{u}^{t+1} \) is computed as
\begin{equation}\label{eq:implicit_fullspace}
\mathbf{u}^{t+1} = \argmin_{\mathbf{u}} \left[
\frac{1}{2\Delta t^2} \| \mathbf{u} - \bar{\mathbf{u}}^{t+1} \|_{\mathbf{M}}^2 + \Psi(\mathbf{u})
\right],
\end{equation}
where \( \bar{\mathbf{u}}^{t+1} = \mathbf{u}^t + \Delta t\, \dot{\mathbf{u}}^t \) is the velocity-predicted configuration, and \( \Psi(\mathbf{u})\) is the total potential energy including work done by external forces.

Directly solving Eq.~\eqref{eq:implicit_fullspace} is expensive for large meshes as the linear system is extremely large. To reduce the computational burden, one introduces a low-dimensional subspace of dimension \( k \ll N \), within which the solution is constrained. Classical approaches approximate the displacement at time $t$ as \( \mathbf{u} \approx \mathbf{B} \mathbf{q} \) using a linear basis \( \mathbf{B} \in \mathbb{R}^{N \times k} \), often obtained via Principal Component Analysis (PCA). However, this approach fails to capture geometric complexity when the solution space lies on a curved nonlinear manifold~\cite{Fulton:LSD:2018}.

To overcome this limitation, a nonlinear map \( f: \Omega \subset \mathbb{R}^k \to \mathbb{R}^N \) is introduced, where \( \mathbf{u} = f(\mathbf{q}) \) parameterizes the high-dimensional displacement through a latent coordinate \( \mathbf{q} \in \Omega \). Substituting this into the original optimization yields:
\begin{equation}\label{eq:implicit_reduced}
\mathbf{q}^{t+1} = \argmin_{\mathbf{q}} \left[
\frac{1}{2\Delta t^2} \| f(\mathbf{q}) - \bar{f}^{t+1} \|_{\mathbf{M}}^2 + \Psi(f(\mathbf{q}))
\right],
\end{equation}
where \( \bar{f}^{t+1} = f(\mathbf{q}^t + \Delta t\, \dot{\mathbf{q}}^t) \) corresponds to the forward prediction in the latent space.

This formulation reduces computational cost while better capturing nonlinear deformation behavior, provided the mapping \( f \) is expressive and efficient to evaluate and differentiate.

\subsection{Learning the Configuration Mapping}
To construct the nonlinear mapping $f$, previous works \cite{Fulton:LSD:2018, Sharp:2023:datafree} propose to parameterize $f$ with a multilayer perceptron (MLP) possessing trainable weights $\theta$.
Let $f_\theta: \mathbb{R}^k \to \mathbb{R}^N$ denote this mapping, and $E_\theta: \mathbb{R}^N \to \mathbb{R}^k$ its encoder counterpart.
Given a collection of observed system configurations
$Q = \{\mathbf{u}_i \in \mathbb{R}^N\}$, the autoencoder framework is employed to learn a compact latent representation that captures the intrinsic structure of the configuration manifold.

The parameters $\theta$ are learned by minimizing the reconstruction loss
\begin{equation}\label{eq:recon_loss}
\min_{\bm{\theta}}
\frac{1}{|\mathcal{B}|}
\sum_{\mathbf{u}_i \in \mathcal{B}}
\| f_\theta(E_\theta(\mathbf{u}_i)) - \mathbf{u}_i \|_M^2,
\end{equation}
where $\mathcal{B}\subset Q$ is a mini-batch of training configurations.
The encoder $E_\theta$ is used only during training to infer latent coordinates that best reconstruct each state $\mathbf{u}_i$, while the decoder $f_\theta$ provides the mapping employed at test time.
The norm $\|\cdot\|_M$ is induced by the manifold metric, ensuring that reconstruction accuracy is measured consistently with the geometric structure of the data.

Minimizing Eq.~\eqref{eq:recon_loss} thus yields a decoder that maps reduced coordinates to the full space, enabling nonlinear model reduction for complex deformation manifolds.

\begin{figure}
    \centering
    \includegraphics[width=\linewidth]{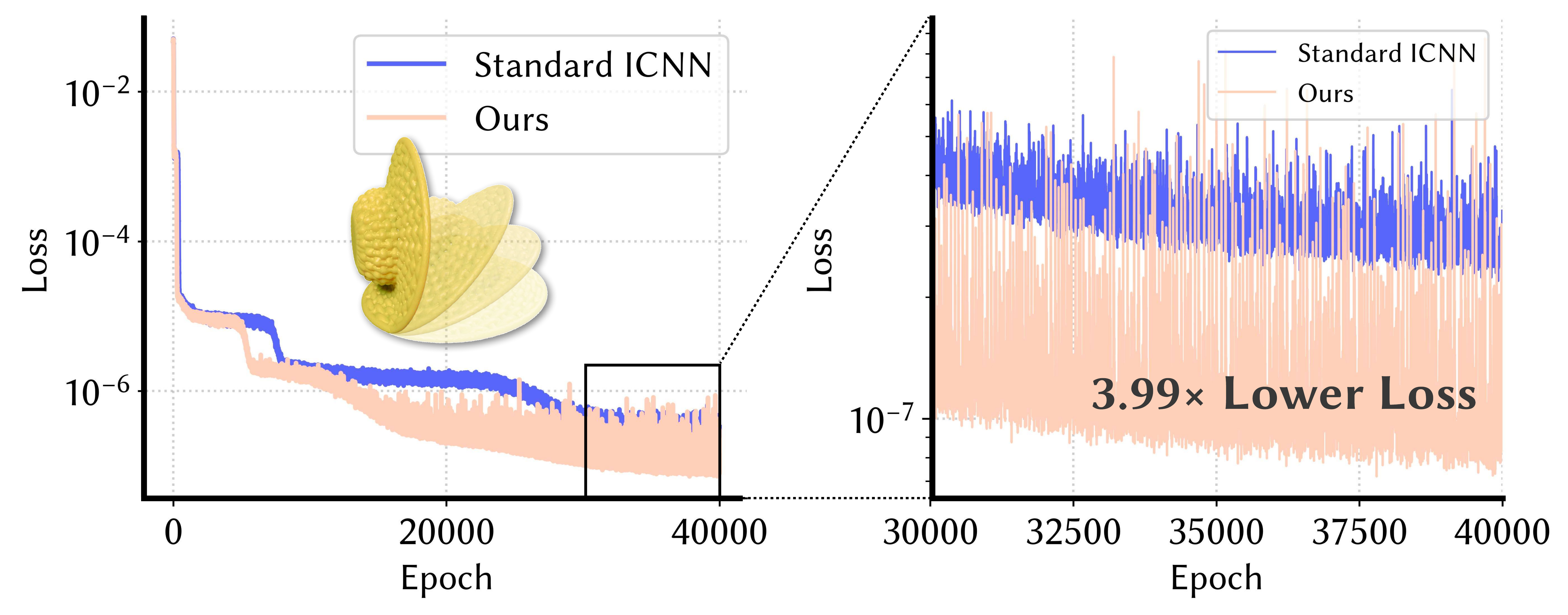}
    \caption{Ablation study comparing convexity regularization in the full space and the reduced space. 
Enforcing convexity in the reduced space achieves faster convergence and lower loss, benefiting from PCA-based initialization. 
In contrast, initialization is not feasible in the full-space formulation due to stricter non-negativity constraints on the weight matrices.}
    \label{fig:loss}
\end{figure}

\section{Regularizing Neural Subspace with Difference-of-Convex Function}

\subsection{Motivation}

In nonlinear model reduction, the displacement field is represented as a nonlinear mapping
$\mathbf{u}=f(\mathbf{q})$ from reduced coordinates $\mathbf{q}\in\mathbb{R}^k$.
Our objective is to design a reduced-order structure that generalizes beyond the training distribution, particularly to larger load magnitudes and reversed force directions, while maintaining a compact latent representation.

Existing approaches typically rely on loss-based regularizers to encourage smoothness or stability in the latent space~\cite{Sharp:2023:datafree, lyu2024accelerate, Liu2022Lipschitz}. Although effective within the sampled domain, such regularization acts only implicitly through the objective function and requires careful tuning against task-specific losses. Moreover, these penalties constrain behavior only at sampled points; since the latent space is unbounded, they provide no guarantee of consistent responses in unsampled regions, where extrapolation is most critical.

We instead embed the desired inductive bias directly into the mapping
$f:\mathbb{R}^k\rightarrow\mathbb{R}^N$. Our formulation adopts an odd-symmetric Difference-of-Convex (odd-DC) structure, expressing $f$ as the difference of two convex functions. Convexity supplies a global regularity prior that discourages oscillations across the entire latent space, improving extrapolation to unseen magnitudes. Antisymmetrizing the convex primitive introduces an odd mapping, enabling consistent responses under inverse loading directions, a property not attainable with convexity alone.

This DC construction relaxes overly restrictive linear assumptions such as PCA while retaining structured smoothness, yielding a nonlinear decoder that responds more coherently on out-of-distribution reduced states. Figure~\ref{fig:overview} summarizes the proposed formulation.

\subsection{Regularizing the Reduced Space}

Rather than imposing structure directly on the full displacement mapping, we apply regularization in an intermediate \emph{reduced space}.  
We define a latent representation
\begin{equation}
\label{eq:reduced_space}
    \mathbf{z} = f(\mathbf{q}), \qquad 
    f : \mathbb{R}^k \rightarrow \mathbb{R}^r,\quad 
    k < r \ll N,
\end{equation}
and reconstruct the displacement through a linear trainable basis
\begin{equation}
    \mathbf{u} = \mathbf{W}\,\mathbf{z}, \qquad \mathbf{W} \in \mathbb{R}^{N\times r}.
\end{equation}
Similar nonlinear--linear staged constructions were explored in~\cite{Fulton:LSD:2018} for their efficiency and stability benefits.  
In our formulation, this separation localizes structural constraints to the latent mapping while leaving the reconstruction layer $\mathbf{W}$ unaffected.  
As shown in Figure~\ref{fig:loss}, this organization also leads to faster convergence in our setting.  
The decoupling allows the nonlinear stage to be regularized without over-restricting the final displacement representation and keeps $\mathbf{W}$ replaceable for later extensions such as neural-field bases.

\subsection{Odd-Symmetric Difference-of-Convex Construction}

To improve generalization to both larger load magnitudes and inverse force directions, we structure the latent mapping using an \emph{odd difference-of-convex} composition,
\begin{equation}\label{eq:odd}
    f(\mathbf{q})
    = f_{\text{convex}}(\mathbf{q})
      - f_{\text{convex}}(-\mathbf{q}),
\end{equation}
which satisfies $f(-\mathbf{q})=-f(\mathbf{q})$ by design.  
The nonconvexity of the overall mapping arises solely from the difference of two convex evaluations, while the only trainable nonlinear component remains the convex primitive $f_{\text{convex}}$.  
This antisymmetrization is a fixed transformation applied at evaluation and therefore does not enlarge the optimization search space.  
As illustrated in Figure~\ref{fig:odd_illus}, the odd-DC formulation extends learned deformations coherently to inverse directions, promoting consistent responses outside the training range.

The above construction requires a parameterization of $f_{\text{convex}}$ that remains convex for all inputs; generic neural networks cannot guarantee this property.  
Consequently, a model class with built-in convexity is essential to realize the proposed structural prior.

\subsection{Input Convex Network as the Convex Primitive}

We construct the convex primitive function $f_{\text{convex}}$ using an Input Convex Neural Network (ICNN)~\cite{amos2017icnn}, which guarantees convexity of the output with respect to the input by design.

\begin{figure}
    \centering
    \includegraphics[width=0.7\linewidth]{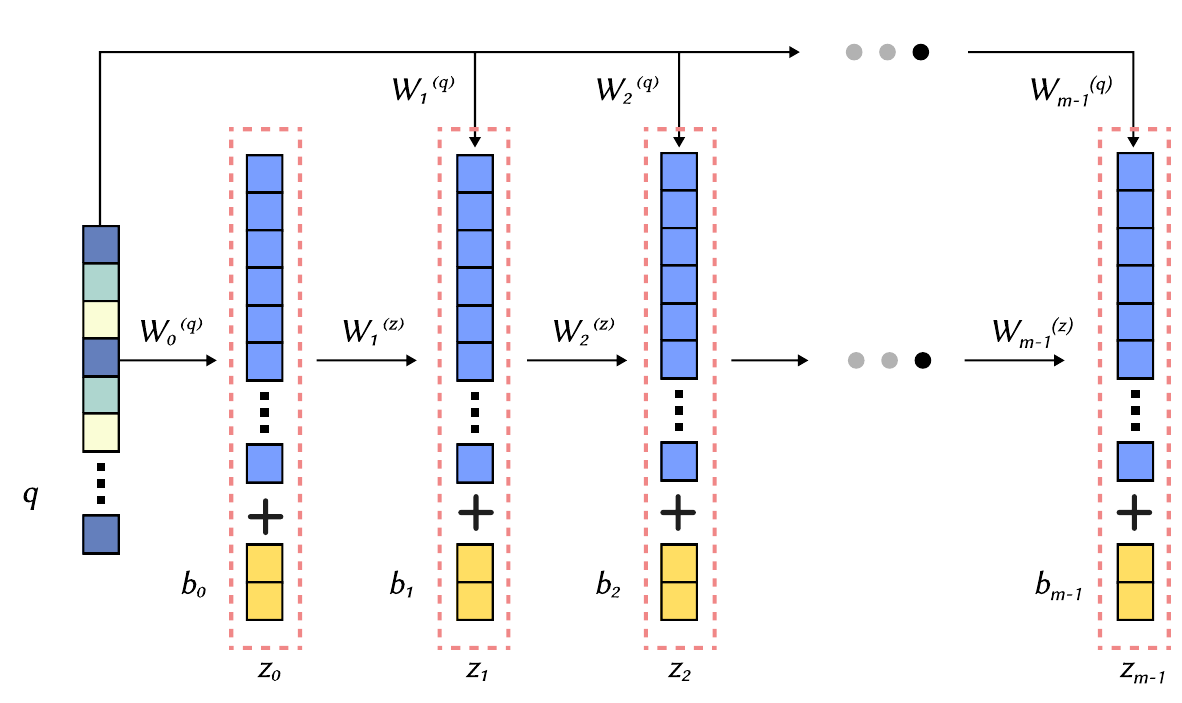}
    \caption{The convex component in odd-DC is built upon the input-convex neural network (ICNN), in which the output is, by construction, convex with respect to the input. 
This convexity is ensured through non-negative weights and non-decreasing activation functions.
}
    \label{fig:icnn_overview}
\end{figure}

As shown in Figure~\ref{fig:icnn_overview}, an ICNN defines hidden activations $\mathbf{h}_0,\dots,\mathbf{h}_m$ via
\begin{equation}\label{eq:icnn_architecture}
    \mathbf{h}_{i+1} = g_i\!\left(\mathbf{W}_i^{(h)}\mathbf{h}_i + \mathbf{W}_i^{(q)}\mathbf{q} + \mathbf{b}_i\right),\quad i = 0,\dots,m{-}1,
\end{equation}
with $f_{\text{convex}}(\mathbf{q}) = \mathbf{h}_m$.  
Convexity is ensured by enforcing $\mathbf{W}_i^{(h)} \!\geq\! 0$ for $i \!\geq\! 1$ and using convex, non-decreasing activation functions (e.g., ReLU or softplus); see~\cite{amos2017icnn} for details.

\begin{figure}
    \centering
    \includegraphics[width=1\linewidth]{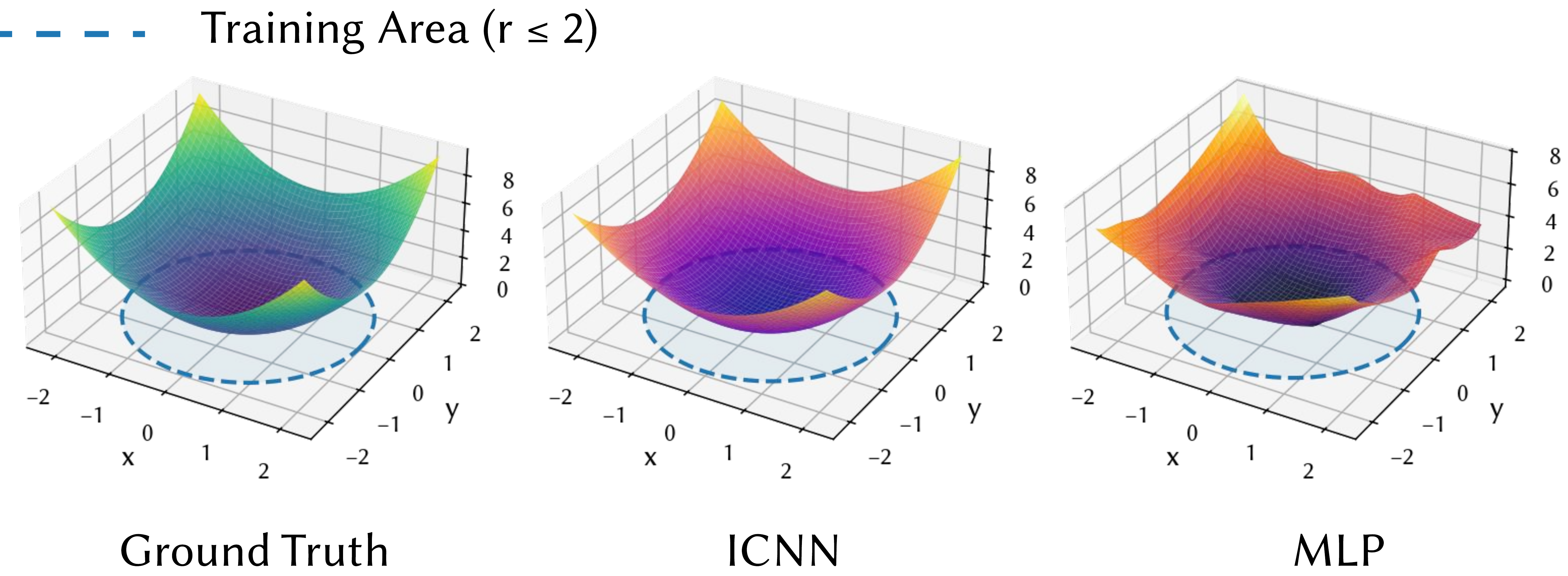}
    \caption{Convexity-enforcing architectures enable smooth extrapolation. An input-convex network and a standard MLP are trained on samples of $z = x^2 + y^2$ within a disk of radius 2. The convex model preserves the global convex structure of the function and generalizes smoothly outside the training region, whereas the unconstrained MLP exhibits non-smooth artifacts when extrapolating.}
    \label{fig:2d_didactic}
\end{figure}

Figure~\ref{fig:2d_didactic} illustrates this property on a didactic 2D example. Both an ICNN and a standard MLP are trained to fit the convex function $z = x^2 + y^2$ using samples within a radius-2 disk. While the MLP exhibits oscillatory behavior outside the training region, the ICNN extrapolates smoothly due to its convex structure.

By imposing convexity only on the latent mapping $\mathbf{q} \mapsto \mathbf{z}$, the ICNN suppresses unwanted oscillations globally, including in regions without training data. The subsequent linear reconstruction via $\mathbf{W}$ (see Equation~\ref{eq:reduced_space}) retains full expressiveness, ensuring that convexity regularizes only the reduced space without limiting the final displacement field.

\subsection{Training}

We adopt a data-driven approach using a dataset of observed configurations
$Q=\{\mathbf{u}_i\in\mathbb{R}^N\}$.  
At inference time, the reduced model consists solely of the decoder mapping
$\mathbf{u}=\mathbf{W}f(\mathbf{q})$.  
During training, however, the latent coordinates $\mathbf{q}$ corresponding to each $\mathbf{u}_i$ are unknown, and an encoder $E:\mathbb{R}^N\!\rightarrow\!\mathbb{R}^k$ is introduced to infer them.  
We therefore train an autoencoder whose decoder is the proposed structured mapping $f$, while the encoder $E$ is used only to obtain training-time estimates $\mathbf{q}_i=E(\mathbf{u}_i)$.  
The form of $E$ depends on the representation and is described separately for mesh-based and continuous reductions.

\subsubsection{Mesh-based Model Reduction}

For mesh-based reductions, the encoder $E$ is parameterized as an MLP that maps the $N$-dimensional displacement vector to a $k$-dimensional latent code.  
Following~\cite{Fulton:LSD:2018}, we initialize the first linear layer of $E$ and the basis $\mathbf{W}$ using PCA.  
The components $E$, $f$, and $\mathbf{W}$ are jointly optimized with the reconstruction loss
\begin{equation}\label{eq:recon_loss_mesh}
\mathcal{L}_{\text{mesh}} = 
\sum_{\mathbf{u}_i \in Q}
\big\| \mathbf{W} f(E(\mathbf{u}_i)) - \mathbf{u}_i \big\|^2 .
\end{equation}

\subsubsection{Continuous Model Reduction}

For continuous model reduction, the mesh discretization may vary between training and inference: the number of vertices $N$ and their ordering are not fixed.  
This prevents parameterizing the encoder $E$ and basis $\mathbf{W}$ as fixed-size matrices.  
Importantly, the dimension of the reduced space remains constant, so the latent mapping
$f(\mathbf{q}):\mathbb{R}^k\!\rightarrow\!\mathbb{R}^r,\; k<r\ll N$
from Eq.~\ref{eq:reduced_space} is still applicable.

To achieve discretization-agnostic reconstruction, we replace the trainable matrix $\mathbf{W}$ with a continuous basis function following~\cite{chang:2023:licrom}.  
The basis is represented as a neural field $\mathbf{W}(\mathbf{x}_0)$ that maps the rest position of any vertex $i$, $\mathbf{x}_0^i$, to a local basis $\mathbf{W}_i\in\mathbb{R}^{d\times r}$, where $d\!\in\!\{2,3\}$ is the spatial dimension.  
Because the latent dimension $r$ is fixed, the full-space displacement for an arbitrary set of vertices can be reconstructed as
\begin{equation}
\label{eq:crom_icnn}
\mathbf{u}(\mathbf{x}_0)=\mathbf{W}(\mathbf{x}_0)\,f(\mathbf{q}), 
\end{equation}
independent of mesh resolution or ordering.

The remaining question is how to construct an encoder compatible with this discretization-agnostic representation.  
Because the basis dimension $r$ is fixed regardless of the number of vertices, the input to the encoder can be expressed in this reduced basis.  
For any configuration $\mathbf{u}$, we first project it using the continuous basis to obtain an $r$-dimensional feature $\mathbf{u}\,\mathbf{W}(\mathbf{x}_0)^{\!\top}$, which is independent of mesh resolution.  
The encoder is then parameterized as a mapping $E:\mathbb{R}^r\!\rightarrow\!\mathbb{R}^k,\; k<r$, and the latent coordinate is computed as $E(\mathbf{u}\,\mathbf{W}(\mathbf{x}_0)^{\top})$.

We optimize the continuous model by minimizing the reconstruction loss over all training configurations and their sampled vertices:
\begin{equation}\label{eq:recon_loss_cont}
\mathcal{L}_{\text{crom}} =
\sum_{\mathbf{u}_i \in Q}
\sum_{\mathbf{x}_0^j \in \mathcal{V}_i}
\big\|
\mathbf{W}(\mathbf{x}_0^j)\,
f\Big( E\big( \mathbf{u}_i\,\mathbf{W}(\mathbf{x}_0^j)^{\!\top}\big)\Big)
-\mathbf{u}_i(\mathbf{x}_0^j)
\big\|^2 ,
\end{equation}
where $\mathcal{V}_i$ denotes the set of sampled rest positions (e.g., vertices) for configuration $\mathbf{u}_i$, and $\mathbf{u}_i(\mathbf{x}_0^j)\in\mathbb{R}^d$ is the displacement at $\mathbf{x}_0^j$.

\begin{figure}
    \centering
    \includegraphics[width=1.0\linewidth]{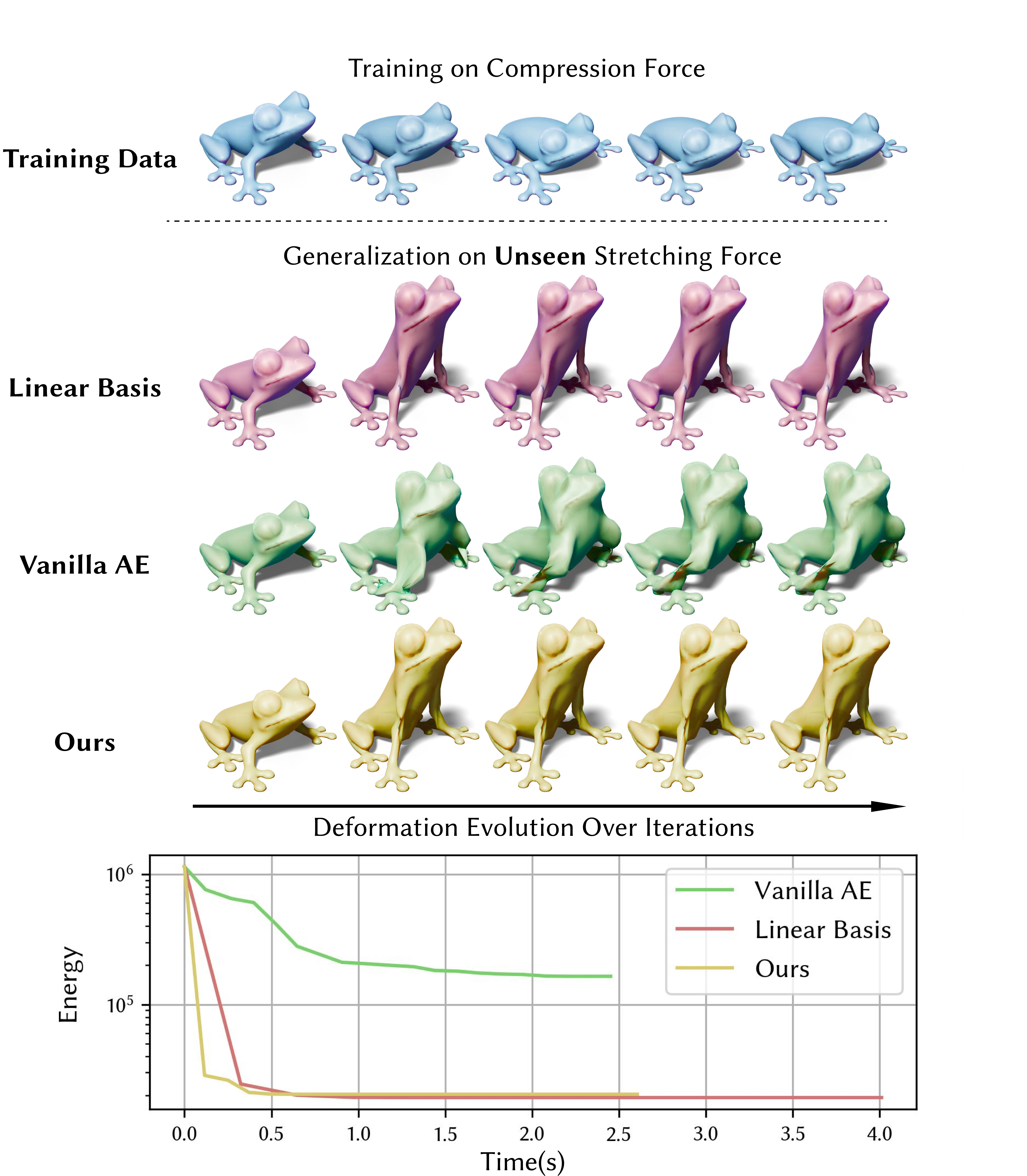}
    \caption{Generalization to unseen force directions. 
We train each model on a compression trajectory (top row) and test on a stretching force that is not included in the training data. 
The vanilla autoencoder fails to produce plausible deformations under stretching, as it lacks structural constraints to guide extrapolation beyond the training range. 
In contrast, our convex-inspired nonlinear formulation captures the correct deformation behavior under the inverted loading direction. 
Thanks to its structured design and compact representation, our method achieves generalization performance comparable to the linear model while converging faster due to the smaller reduced system.
}
    \label{fig:generalization_loading}
\end{figure}
\begin{figure}
    \centering
    \includegraphics[width=1.0\linewidth]{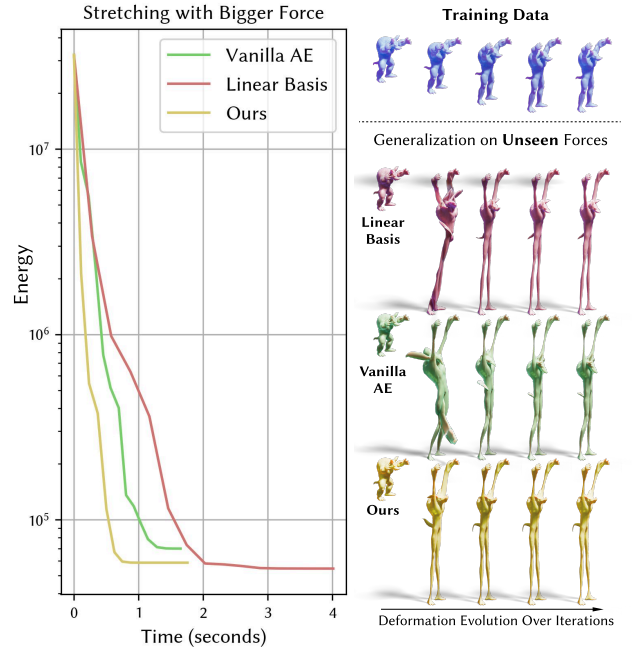}
    \caption{Generalization to unseen force magnitudes. 
As shown on the right, the model is trained under a smaller stretching force. 
During testing, we apply a larger unseen stretching force and minimize the elastic energy to obtain the steady-state deformation. 
Thanks to the structured nonlinear mapping, our method achieves the best convergence and most stable steady states among all baselines, as shown on the left.}
    \label{fig:generalization_loading2}
\end{figure}

\begin{table}[ht]
\centering
\renewcommand{\arraystretch}{1.15}
\colorlet{rowgray}{gray!15}
\begin{minipage}{\linewidth}
\centering
\begin{tblr}{
  colspec = {Q[l,wd=1.4cm] Q[c,wd=1.0cm] | c c c},
  column{1} = {leftsep=2pt,rightsep=2pt, font=\footnotesize},
  column{2} = {leftsep=2pt,rightsep=2pt},
  vline{3} = {0.6pt},
  hline{1,Z} = {0.8pt},
  hline{2} = {3-5}{0.6pt},
  hline{3} = {0.6pt},
  cell{1}{2} = {r=2}{c,m},
  cell{1}{3} = {c=3}{c,m},
  row{4,6,8,10,12} = {bg=gray!15},
}
\textbf{} & \textbf{Latent\\dim.} & \textbf{Reconstruction Error ($\downarrow$)} & & \\
\textbf{} & & \textbf{Vanilla AE} & \textbf{Ours} & \textbf{Linear Basis} \\
{\textbf{Cute Dragon}} & 8 & $1.46\times10^{-7}$ & $3.62\times10^{-7}$ & $1.27\times10^{-5}$ \\
\textbf{Xyz Dragon}  & 8 & $2.60\times10^{-8}$ & $7.90\times10^{-8}$ & $2.28\times10^{-6}$ \\
\textbf{2D Bar}      & 2 & $7.51\times10^{-7}$ & $2.75\times10^{-6}$ & $9.36\times10^{-6}$ \\
\textbf{Unicorn}     & 8 & $4.14\times10^{-4}$ & $2.22\times10^{-4}$ & $4.78\times10^{-5}$ \\
\textbf{Hand}        & 6 & $5.11\times10^{-8}$ & $1.40\times10^{-7}$ & $3.00\times10^{-7}$ \\
\textbf{Frog}        & 8 & $3.12\times10^{-8}$ & $8.08\times10^{-8}$ & $8.76\times10^{-9}$ \\
\textbf{Armadillo}   & 8 & $1.46\times10^{-7}$ & $3.62\times10^{-7}$ & $1.27\times10^{-5}$ \\
\textbf{Boat}        & 3 & $3.16\times10^{-4}$ & $1.70\times10^{-3}$ & $5.38\times10^{-2}$ \\
\textbf{Whale}       & 8 & $3.49\times10^{-8}$ & $7.84\times10^{-8}$ & $8.33\times10^{-7}$ 
\end{tblr}
\end{minipage}
\caption{
Reconstruction error comparison under identical latent dimensions.
All methods in this table are evaluated using the same latent dimension to assess reconstruction capacity under a fixed-dimensional setting.
In the full experimental pipeline, the latent dimension of the linear basis is increased as needed to match the reconstruction accuracy of nonlinear methods.
Under this constraint, our method achieves reconstruction accuracy comparable to the vanilla autoencoder while outperforming the linear basis in most cases.}
\label{tab:recon_error}
\end{table}


\begin{table}[t]
\centering
\rowcolors{2}{white}{gray!20}
\begin{tabularx}{\linewidth}{
l |
>{\centering\arraybackslash}X
>{\centering\arraybackslash}X
>{\centering\arraybackslash}X
|>{\centering\arraybackslash}X}
 & \scriptsize{\textbf{Unseen compression}} 
 & \scriptsize{\textbf{Reconstruction of Training}} 
 & \scriptsize{\textbf{Larger stretching}}
 & \footnotesize{\textbf{Average Conv. time}} \\
\scriptsize{\textbf{Vanilla AE}}      & 22.38\%  & 1.67\%  & 4.80\% & 0.1729s\\
\scriptsize{\textbf{Linear Basis}}    & 13.51\%  & 1.75\%  & 3.31\% & 0.3258s \\
\scriptsize{\textbf{Ours}}            & 14.07\%  & 1.69\%  & 3.26\% & 0.1296s \\
\end{tabularx}
\caption{Average percentage difference from the ground-truth deformation and the convergence time, aggregated across three evaluation scenarios and over more than 50 test cases. Across these scenarios, our method consistently exhibits lower deviation than the vanilla autoencoder, while achieving error levels comparable to the linear basis, which similarly preserves the underlying deformation structure.
In addition, our method converges fastest on average across the 50+ test cases.}
\label{tab:generalization_error}
\end{table}

\section{Results}
We evaluate our method across a range of settings. All networks are trained with Adam~\cite{kingma2014adam} for 50k epochs at a learning rate of $10^{-4}$. For mesh-based models, the first and last layers are initialized using PCA. Unless otherwise specified, training data are generated with a full-space FEM solver using stable Neo-Hookean elasticity~\cite{Smith:2018:stablenh}.



\subsection{Generalization on Unseen Loading}

\begin{figure}
    \centering
    \includegraphics[width=0.8\linewidth]{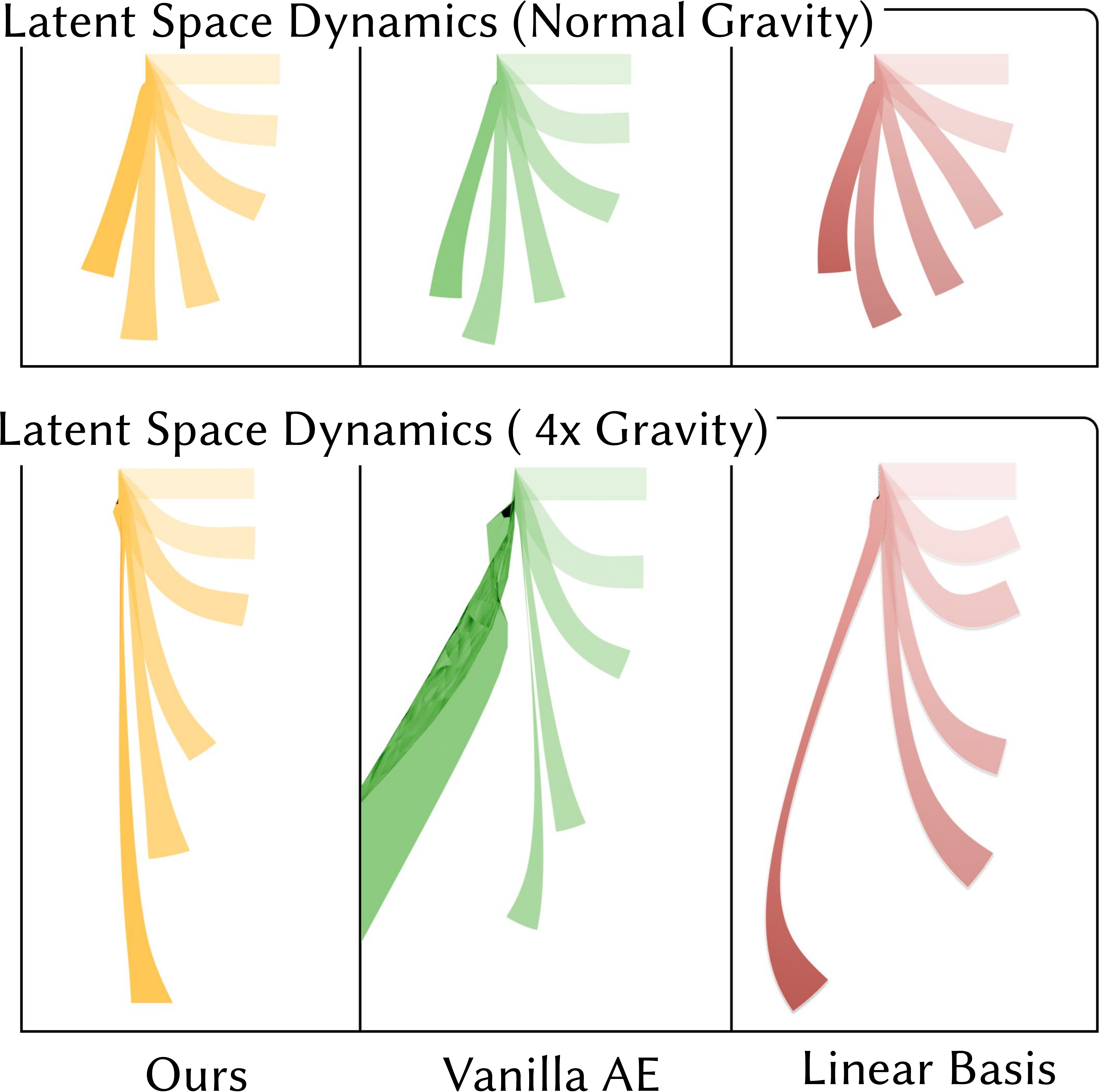}
    \caption{Latent-space dynamics under increased loading magnitude. Rollouts of a 2D bar under normal gravity (training regime) and under four-times increased gravity (out-of-distribution). While all methods exhibit stable dynamics under normal gravity, the vanilla autoencoder becomes unstable under stronger loading, producing non-physical behavior. In contrast, both the linear basis and the proposed structured formulation maintain stable and coherent deformation dynamics when extrapolating to larger-magnitude loading.}
    \label{fig:generalization_loading3}
\end{figure}

We evaluate our model’s generalization capability under two out-of-distribution (OOD) scenarios, \textbf{loading direction} and \textbf{loading magnitude}, using three representative cases: opposite-direction loading on the \textit{Frog} (Fig.~\ref{fig:generalization_loading}), larger-magnitude loading on the \textit{Armadillo} (Fig.~\ref{fig:generalization_loading2}), and increased-gravity loading on a 2D bar (Fig.~\ref{fig:generalization_loading3}). In all cases, models are trained within a limited loading regime and evaluated beyond the training distribution to assess robustness.

For the \textit{Frog} (Fig.~\ref{fig:generalization_loading}), the training data consist solely of compressive deformations generated by downward loading. At test time, we apply an equal-magnitude tensile load in the upward direction, implemented via fixed-point constraints. Starting from the rest state, we directly iterate under this unseen loading condition. We evaluate generalization by tracking the potential energy across Newton iterations, using energy decay as a measure of convergence.

This setup explicitly probes the model’s ability to extrapolate beyond the training regime and recover physically consistent behavior under reversed loading conditions. The linear basis model reduction preserves a coherent, physically meaningful stretching pattern across iterations. The vanilla autoencoder (AE) remains numerically stable but fails to produce a plausible stretching deformation, often generating distorted shapes inconsistent with expected tensile behavior. In contrast, our symmetric odd-function formulation achieves shape quality comparable to the linear basis, faithfully capturing the intended stretch. Moreover, it converges to a low-energy configuration faster, highlighting its improved generalization under unseen loading directions.

For the \textit{Armadillo} (Fig.~\ref{fig:generalization_loading2}), the training dataset covers moderate loading magnitudes, while the test scenario applies substantially stronger forces that exceed the training range. This experiment evaluates the framework’s ability to generalize to high-strain regimes where nonlinear effects dominate. As in Fig.~\ref{fig:generalization_loading}, we implement fixed-point constraints and iterate directly from the rest state via energy minimization. As shown in Fig.~\ref{fig:generalization_loading2}, our symmetric and convex formulation achieves markedly faster energy reduction than the linear baseline, while the vanilla AE plateaus at a higher energy level and fails to reach the minimal state. These results highlight our formulation’s efficiency and robustness when extrapolating to unseen, high-magnitude loading conditions.

This improved generalization also enables more stable simulation. For the 2D bar (Fig.~\ref{fig:generalization_loading3}), the training data are generated under normal gravity, under which all methods produce stable and physically plausible rollouts in the latent-space dynamics. When the gravity magnitude is increased by a factor of four—exceeding the training range—the same simulation procedure reveals markedly different behavior. Under this stronger loading, both the linear basis and our proposed method remain stable and preserve coherent deformation behavior, whereas the vanilla AE becomes unstable during the rollout, exhibiting non-physical deformations. This result indicates that, when extrapolating to larger-magnitude loading, enforcing structure in the latent mapping substantially improves robustness.


\subsection{Performance Comparison on 50+ Evaluation Cases}

\begin{figure*}
    \centering
    \includegraphics[width=1.0\linewidth]{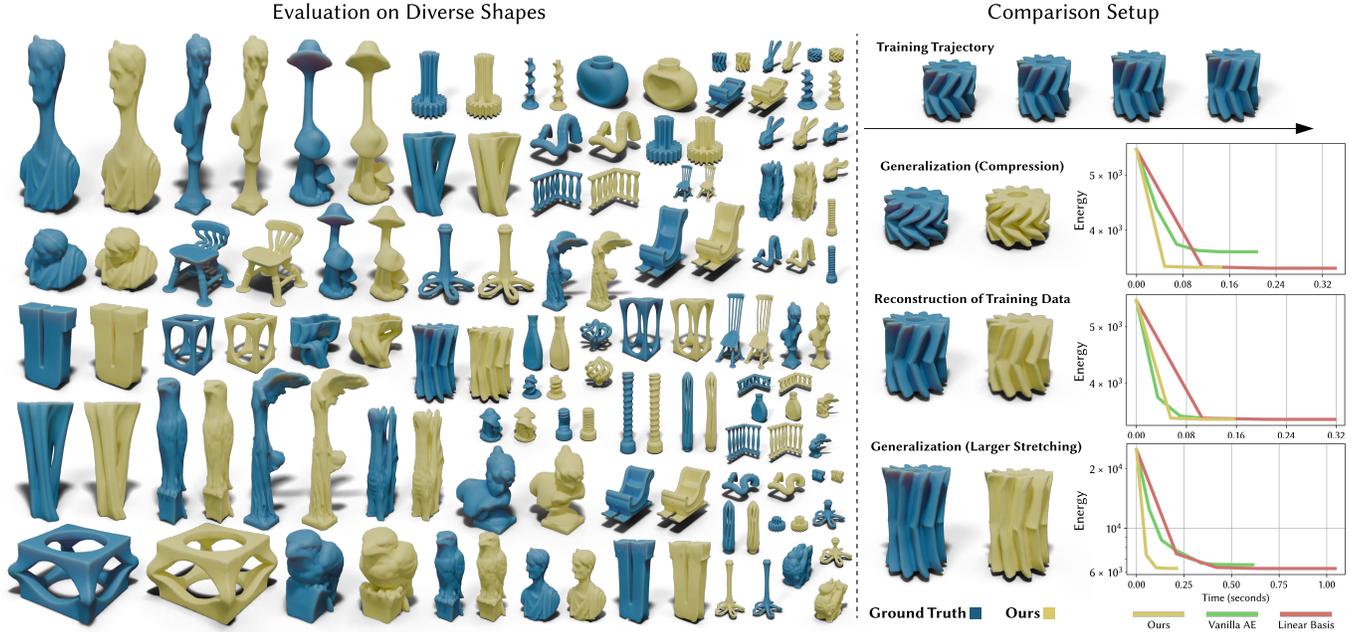}
    \caption{
    Evaluation on a diverse set of over 50 test cases.
    The left panel shows the collection of shapes used for testing.
    The right panel illustrates the comparison setup: models are trained on a single deformation trajectory (top), and then evaluated under three settings: generalization to unseen compression, reconstruction of the training data, and generalization to larger stretching.
    This evaluation protocol is applied consistently across all test cases.
    }

    \label{fig:mini_dataset}
\end{figure*}

We also evaluate our method on a collection of 50+ test cases (Figure~\ref{fig:mini_dataset}). The evaluation set includes multiple object instances, each subjected to several distinct deformation scenarios. For each object, the model is trained on a single deformation trajectory and then evaluated on that same loading as well as two additional, unseen loadings. These loadings are implemented using fixed-point constraints, and Newton's method is used to minimize the elastic energy for a single static frame. Performance is measured by comparing the resulting deformed configurations against the corresponding ground-truth states.

Across all test cases, our method consistently achieves accurate matching on the two unseen deformations. It outperforms the vanilla autoencoder on both unseen cases and shows comparable performance to the linear basis on one case, while outperforming it on the other (Table~\ref{tab:generalization_error}). Although linear bases perform well for certain unseen deformations due to their fixed structured formulation, they require substantially higher reduced dimensions and consequently incur higher computational cost. In contrast, our method achieves comparable or better accuracy on unseen deformations using a more compact reduced representation, resulting in nearly a twofold reduction in computation time. For the training deformation, all methods accurately match the final configuration, with the vanilla autoencoder occasionally achieving lower error due to its unconstrained expressiveness on the observed data. Overall, our method provides a more favorable trade-off between accuracy, generalization robustness, and computational efficiency across deformation scenarios.

\subsection{Comparison with Other Regularization Methods} 

\begin{figure}
    \centering
    \includegraphics[width=1.0\linewidth]{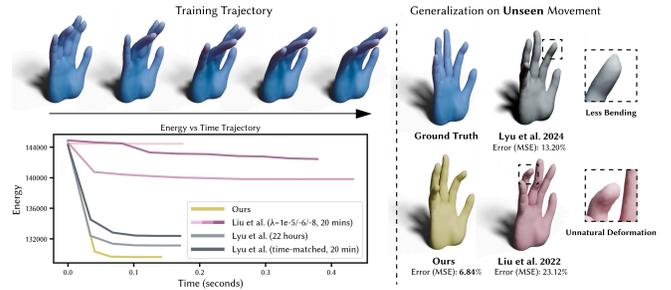}
\caption{
Comparison of neural regularization methods on the hand model. All methods are trained on the same hand motion trajectory (top left) and evaluated on an unseen movement. Our method achieves the lowest final energy and reconstruction error while preserving natural and physically plausible deformations. In contrast, prior regularization approaches~\cite{Liu2022Lipschitz,lyu2024accelerate} neither converge to the lowest energy state nor maintain deformation quality under extrapolation, resulting in higher error and visible artifacts.
}
    \label{fig:hand_comparison_regularization}
\end{figure}

We further compare our method with two existing regularization methods~\cite{lyu2024accelerate, Liu2022Lipschitz}. All methods are trained on the same deformation trajectory of the hand shape and evaluated by applying an unseen movement.

In this comparison (Fig.~\ref{fig:hand_comparison_regularization}), our method achieves the lowest final energy and the smallest mean squared error (MSE) in the resulting deformation, outperforming both existing regularization baselines within the simulation setting. While both baseline methods aim to regularize nonlinear models via Lipschitz constraints, they differ in formulation and were originally developed for different tasks. In particular, the Lipschitz network (LipMLP)~\cite{Liu2022Lipschitz} was designed for neural SDF representation, where it performs well. However, when applied to our deformation task, LipMLP tends to produce less natural results and shows higher sensitivity to hyperparameter choices. This highlights the importance of tailoring regularization strategies to the structure and requirements of physical simulation.

The accelerated Lipschitz network (AccLip)~\cite{lyu2024accelerate} did the regularization by incorporating Hessian-based energy information into the training loss, resulting in improved performance within the training distribution. However, it still shows noticeable deviations from the ground truth in certain regions—such as reduced bending in some fingers—and incurs substantially higher training costs, requiring approximately 22 hours under the same training setup due to explicit evaluation of Hessians with respect to the latent variables. 

Under time-matched constraints (e.g., 20 minutes). While our method and AccLip both target physically informed latent-space models, they pursue different goals: AccLip focuses on accelerating convergence within the training regime, whereas our approach emphasizes generalization to unseen loadings. We view these directions as complementary and potentially synergistic.

\subsection{Optimization with Small Amount of Cubature Points} 
Since our formulation remains nonlinear, it retains the advantage of representing deformation data within a more compact subspace. Consequently, the same amount of deformation can be captured with a lower-dimensional reduced space. This compactness reduces the number of cubature points \cite{An:Cubature:2008} required during optimization, as the number of samples must theoretically exceed the reduced-space dimension to avoid introducing null-space components.

To examine this property, we trained our nonlinear model on a simulated deformation trajectory of a whale mesh and tested its behavior under varying cubature sparsity. Starting from random initial deformations in the latent space, we minimized the elastic energy using one to five randomly selected cubature tetrahedra. The resulting full-space displacement norms after optimization were used to assess reconstruction accuracy. For the linear baseline, we employed a least-squares solve to eliminate null-space components, which ideally projects the system back to its rest pose and yields a zero final displacement norm. To account for randomness in cubature selection, each configuration was optimized 100 times for every cubature count, and the mean and variance of the final displacement norms were recorded.

\begin{figure}
    \centering
    \includegraphics[width=1.0\linewidth]{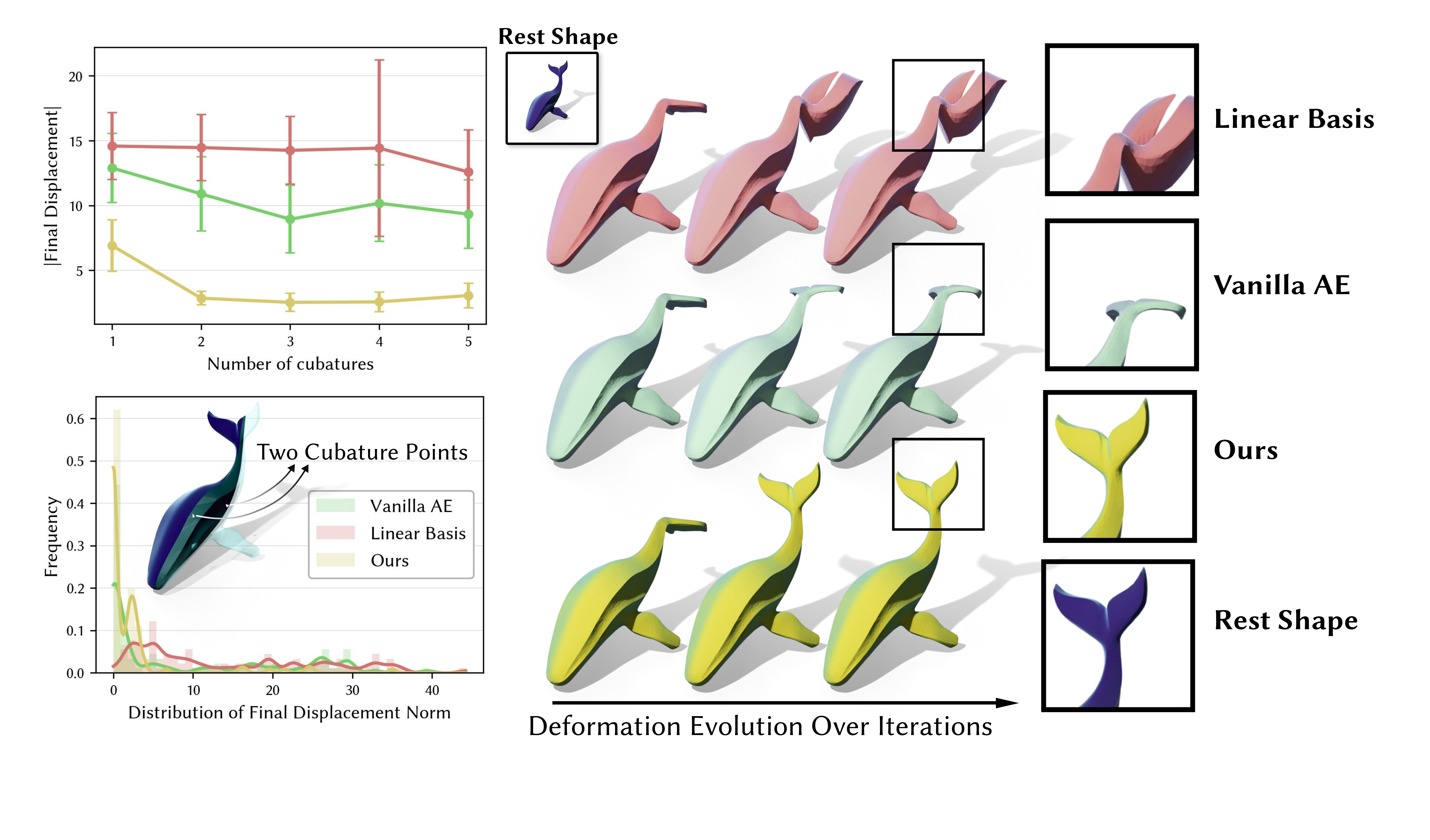}
    \caption{Thanks to its compact representation, our formulation remains stable when using only a few cubature points. 
We train on a simulated whale mesh and minimize elastic energy from random reduced-space initializations with 1–5 cubature tetrahedra. 
As shown (top left), our method yields the smallest final displacement norm and variance across 100 random trials. 
The distribution under two cubature points (bottom left) shows consistently smaller errors, and the visual results (right) demonstrate more natural convergence compared to linear and vanilla autoencoder baselines.}
    \label{fig:robustness_low_cubature}
\end{figure}

As shown in Figure~\ref{fig:robustness_low_cubature}, both nonlinear mappings (the vanilla autoencoder and our symmetric convex formulation) achieve smaller final displacements than the linear model, confirming that nonlinear representations better capture deformation subspaces compactly. Among all three methods, our approach consistently exhibits the smallest average displacement and variance, demonstrating superior robustness and stability even with extremely few cubature samples.

\begin{figure*}
    \centering
    \includegraphics[width=1.0\linewidth]{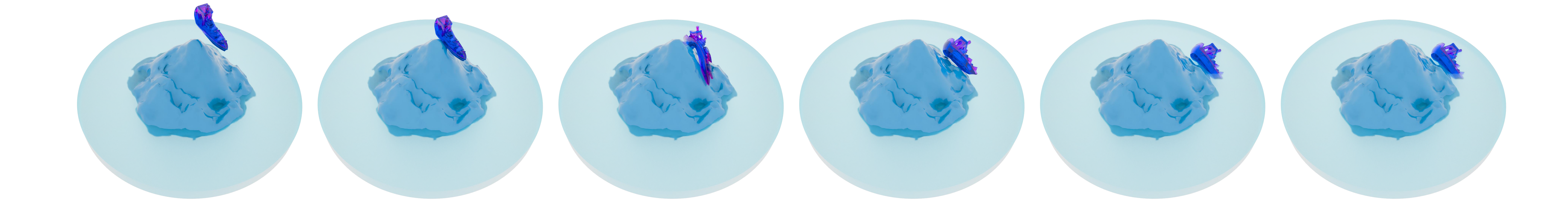}
    \caption{Collision handling. 
Similar to previous methods, our approach supports collision response within reduced-space simulation. 
We run reduced-space dynamics using the nonlinear basis learned by our method. 
In this example, a deformable boat collides with a static mountain and bounces back, demonstrating that our formulation preserves stable and realistic contact behavior.}
    \label{fig:Collision}
\end{figure*}


\subsection{Real-time Simulations} 

\begin{figure}
    \centering
    \includegraphics[width=1.0\linewidth]{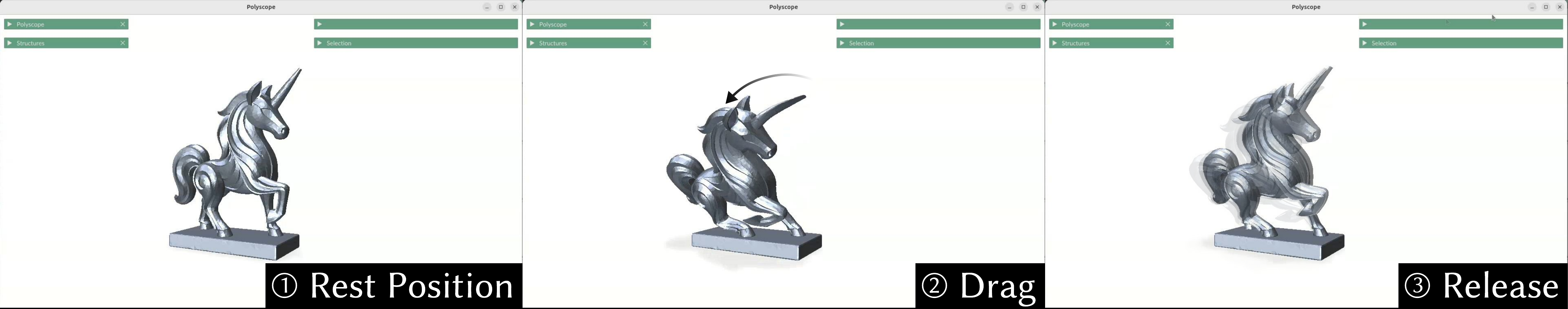}
    \caption{Our method is fast and robust enough to support real-time interaction. 
We demonstrate this through a screen recording of mouse-based manipulation of a deformable unicorn model. 
The unicorn deforms in response to dragging and returns to its rest pose upon release, illustrating stable and responsive real-time behavior.}
    \label{fig:interactive}
\end{figure}

Our method supports real-time interactive applications. 
In this demonstration, a deformable unicorn model is simulated in real time and manipulated through direct mouse interaction. 
The user drags different parts of the model, and the unicorn deforms smoothly and stably in response to the applied forces. 
Upon release, the model returns naturally to its rest configuration, illustrating both the speed and robustness of our reduced nonlinear formulation for interactive physics-based simulation.

Our method also generalizes to scenarios involving collisions. As shown in Figure~\ref{fig:Collision}, we simulate a deformable boat interacting with a static mountain surface. During reduced-space simulation, the learned nonlinear mapping effectively captures the complex contact-induced deformation. Collision response is implemented using the signed distance field (SDF) of the mountain to correct vertex penetrations.

\subsection{Discretization-agnostic Extension} 

\begin{figure}
    \centering
    \includegraphics[width=1.0\linewidth]{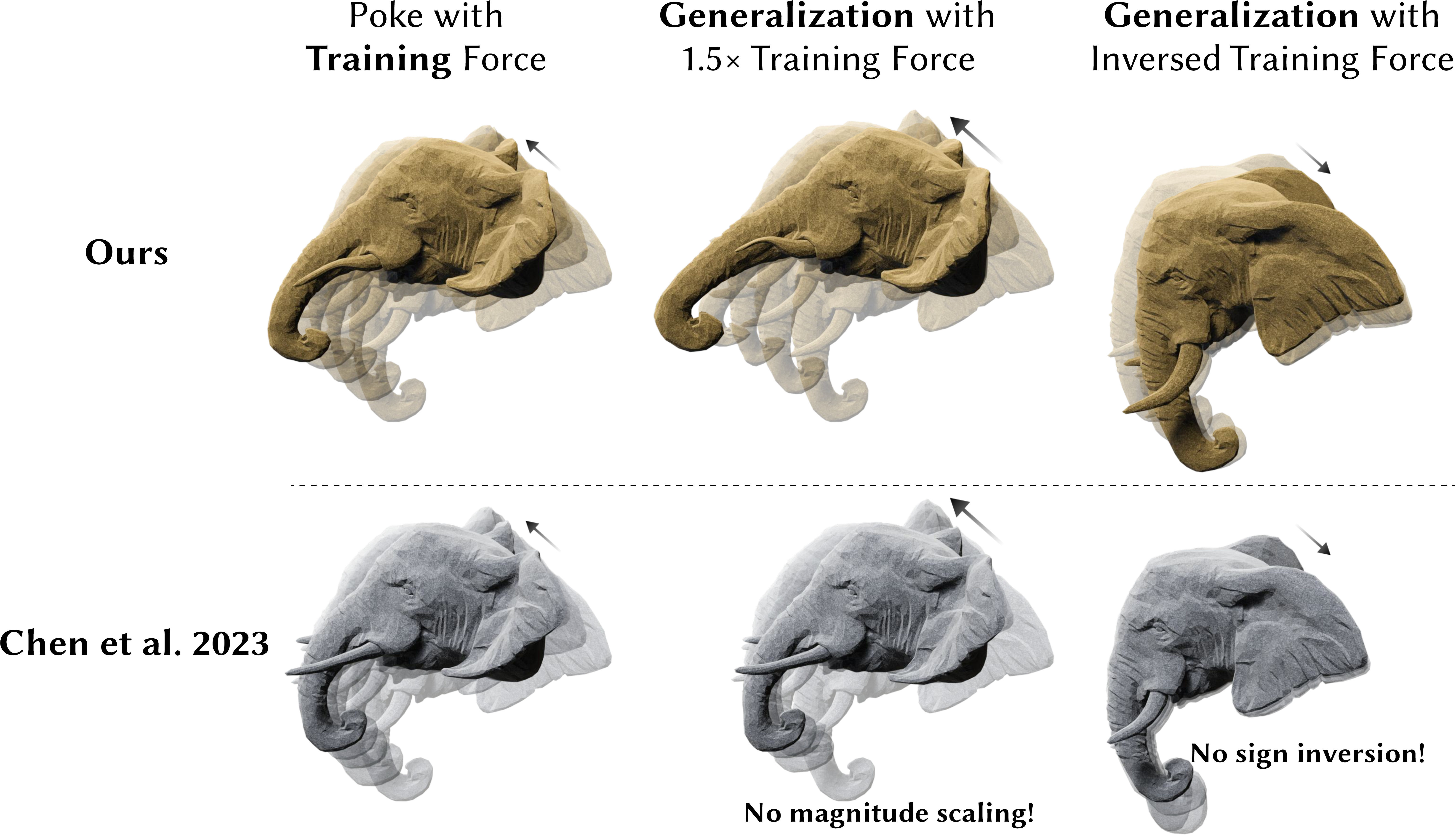}
    \caption{Generalization under unseen loading in continuous model reduction. We compare our regularized formulation (top) with CROM~\cite{chen2023crom} (bottom), a nonlinear reduced model based on implicit neural representations. All models are trained on a single poking force (left column), and tested under 1.5$\times$ force (middle) and inverted force (right). While CROM performs well in-distribution, it fails to generalize to out-of-distribution forces—showing no magnitude scaling or sign inversion. Our method extrapolates correctly, producing plausible deformations across all test cases.}

    \label{fig:elephant_compare_crom}
\end{figure}

We further demonstrate that our regularization method extends beyond mesh-based model reduction to continuous nonlinear settings. Specifically, we apply it to the CROM framework~\cite{chen2023crom}, a discretization-agnostic reduced model based on implicit neural representations. As shown in Figure~\ref{fig:elephant_compare_crom}, we train a model to simulate deformations of an elephant head under a single poking force. We then evaluate generalization under scaled and inverted forces. While CROM performs well within the training regime, it struggles to extrapolate: it fails to scale with increased force magnitude and cannot invert the deformation under force reversal. In contrast, our method generalizes to both scenarios, producing physically plausible responses under unseen loading conditions.

To demonstrate that our method preserves the discretization-agnostic property of prior continuous model reduction approaches, we evaluate a single trained model on meshes of three different resolutions: 2.5K, 50K, and 250K vertices. Despite the varying discretization, the model produces qualitatively similar deformation behavior across all resolutions, indicating robust generalization to unseen mesh discretizations.

\begin{figure}
    \centering
    \includegraphics[width=1.0\linewidth]{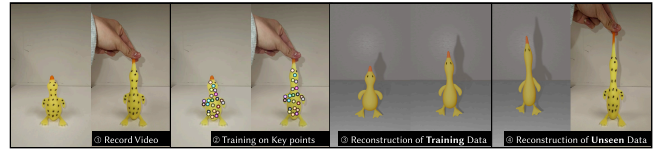}
    \caption{Real-world deformation reconstruction using our continuous formulation and sparse 2D keypoint supervision. (1) A video of a rubber toy undergoing deformation is recorded. (2) A sparse set of 2D keypoints is annotated for a small number of frames. (3) The model is trained to reconstruct full-object deformations from these annotations. (4) Our method generalizes to reconstruct plausible deformations under unseen forces, despite being trained only on limited keypoint data.}

    \label{fig:duck_video}
\end{figure}

Lastly, we demonstrate that the continuous formulation of our method can reconstruct plausible deformations from sparse, real-world observations. We train a deformation model on a rubber toy using only 2D keypoint annotations extracted from a short video sequence (Fig.~\ref{fig:duck_video}). Despite the limited supervision, the model accurately reconstructs full-object deformations and generalizes to new frames involving unseen forces. This example highlights the potential of our formulation and regularization in real-world settings, supporting data-efficient deployment without requiring dense input.

\begin{figure}
    \centering
    \includegraphics[width=1.0\linewidth]{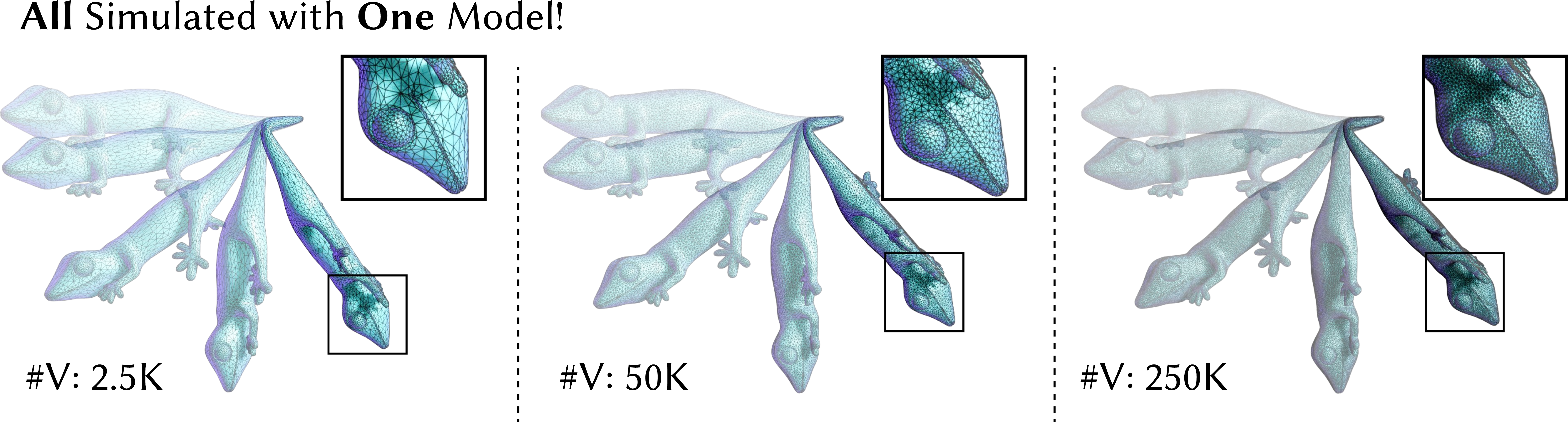}
    \caption{Discretization-agnostic generalization. A single model, trained on a continuous representation, is evaluated on three meshes with increasing resolution (\#V: 2.5K, 50K, 250K). The resulting deformations remain consistent across discretizations, demonstrating that our method inherits the resolution-independence of prior continuous model reduction techniques.}
    \label{fig:lizard_agnostic}
\end{figure}

\section{Discussions and Future Work}
We demonstrate the main advantage of our method through its improved generalization to unseen loading conditions, including both increased force magnitudes and reversed directions. This capability stems from our structured decoder, which integrates convexity and odd symmetry to better capture the physical behavior of deformable systems under out-of-distribution forces.

Due to the compactness of the nonlinear representation, our method achieves comparable deformation accuracy using significantly fewer basis functions, resulting in smaller reduced systems and faster solves compared to traditional linear reduction.

\begin{figure}
    \centering
    \includegraphics[width=1\linewidth]{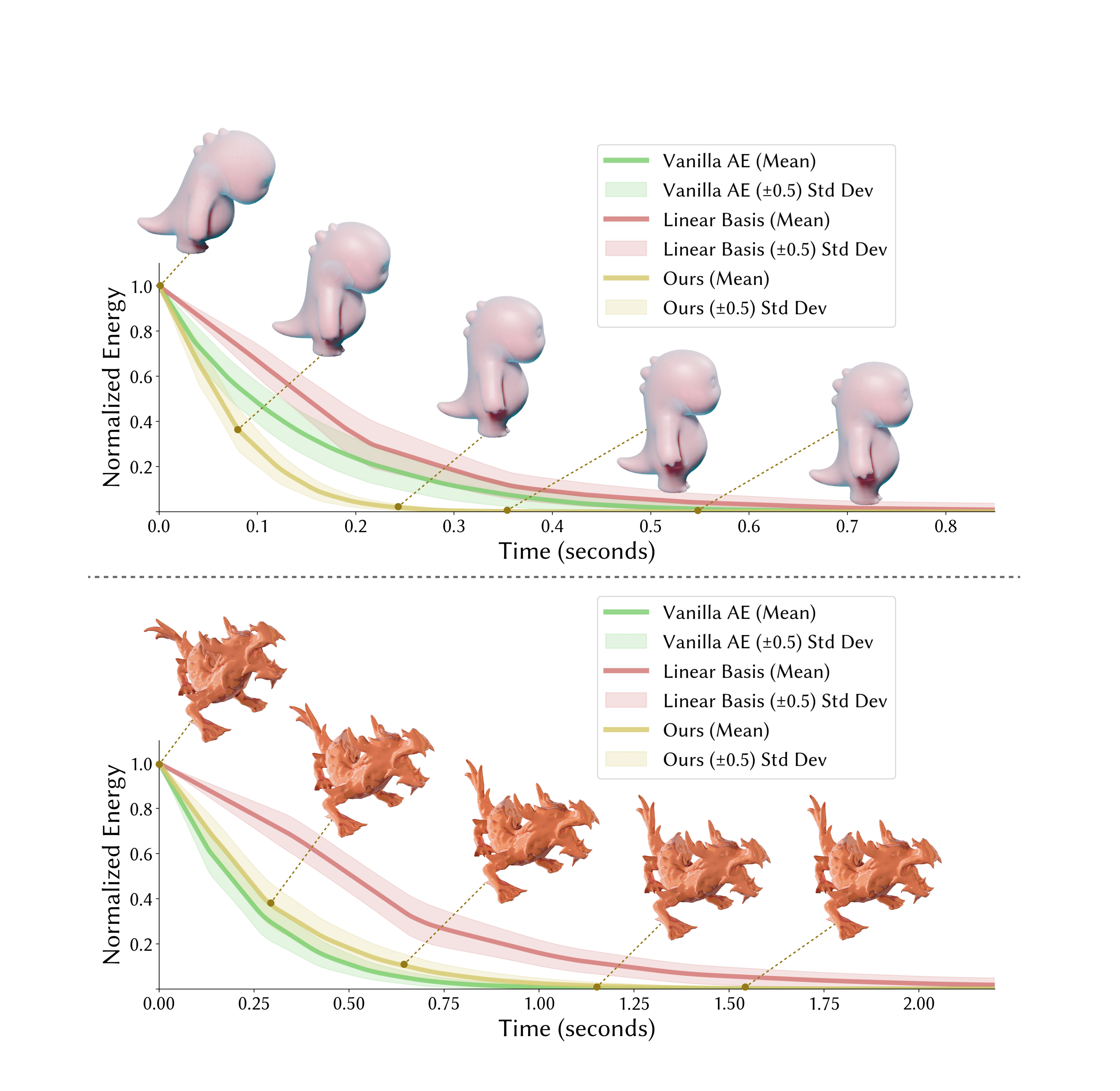}
    \caption{We evaluate convergence behavior on configurations sampled from the training latent codes by minimizing elastic energy from random initializations. 
The plots show the average elastic energy and its variance over optimization time. 
Thanks to its compact representation, our method achieves faster convergence than the linear subspace model. 
Compared to the vanilla autoencoder, convergence is faster in some scenarios (top) and slightly slower in others (bottom), suggesting that convergence behavior may depend on the underlying deformation scenario.
}
    \label{fig:convergence}
\end{figure}

Within regions well covered by training data, our formulation exhibits convergence behavior similar to that of a vanilla autoencoder—converging faster in some cases and slightly slower in others. As shown in Figure~\ref{fig:convergence}, we evaluate convergence by minimizing elastic energy from various starting points in the reduced space and measure progress via normalized energy and its variance over iterations.

Our current approach models only the displacement field and does not explicitly guarantee the absence of local minima in the energy landscape. Certain deformation families, such as those exhibiting bistability, inherently contain multiple stable states that lie outside the scope of our current formulation. In principle, the additional odd-symmetric term can mitigate convexity-induced single-minimum behavior, although we have not observed any degradation in performance during training or simulation. 

Looking forward, several extensions appear promising. Since input-convex neural networks can also represent energy functions, incorporating them into the energy formulation itself could unify model reduction with learned elastic energies. Another direction is to extend our framework to handle multi-stable and highly nonlinear systems, where symmetry-aware convexity might help stabilize transitions between different energy wells. Beyond deformation simulation, the proposed symmetric convex mapping could also benefit related domains such as dynamics prediction, parameter-space exploration, and differentiable design, providing a foundation for more generalizable reduced-order physics models.

\bibliographystyle{ACM-Reference-Format}
\bibliography{main}

\end{document}